\documentclass[10pt,twocolumn,oneside,final]{IEEEtran}
\usepackage{graphicx,cite,amsmath,amssymb,hhline,textcomp}
\begin{document}
\date{}
\title{Performance Analysis of Spectrum Sensing With Multiple Status Changes in Primary User Traffic}
\author{

\medskip {\normalsize $\mbox{Liang Tang}^{}$,
    $\mbox{Yunfei Chen}^{}$, {\em Senior Member, IEEE},
    $\mbox{Evor L. Hines}^{}$, 
    and $\mbox{Mohamed-Slim Alouini}^{}$, {\em Fellow, IEEE}}
    
\IEEEcompsocitemizethanks{\IEEEcompsocthanksitem Liang Tang, Yunfei Chen, Evor L. Hines are with the School of Engineering, University of Warwick, Coventry, CV4 7AL, UK. (e-mail: \{liang.tang,~yunfei.chen,~e.l.hines\}@warwick.ac.uk). 
\IEEEcompsocthanksitem Mohamed-Slim Alouini is with the King Abdullah University of Science and Technology, Thuwal, Makkah Province, Saudi Arabia. (e-mail: slim.alouini@kaust.edu.sa).\protect  
\IEEEcompsocthanksitem This work was supported in part by EPSRC First Grant and in part by Qatar National Research Fund.}}

\IEEEcompsoctitleabstractindextext{%
\begin{abstract}
\boldmath
In this letter, the impact of primary user traffic with multiple status changes on the spectrum sensing performance is analyzed. Closed-form expressions for the probabilities of false alarm and detection are derived. Numerical results show that the multiple status changes of the primary user cause considerable degradation in the sensing performance. This degradation depends on the number of changes, the primary user traffic model, the primary user traffic intensity and the signal-to-noise ratio of the received signal. Numerical results also show that the amount of degradation decreases when the number of changes increases, and converges to a minimum sensing performance due to the limited sensing period and primary holding time.
\end{abstract}

}

\maketitle
\IEEEdisplaynotcompsoctitleabstractindextext   
\IEEEpeerreviewmaketitle

%
%
%
%
%
%
%
%
%
\section{Introduction}
Cognitive radio is a promising solution to the spectrum scarcity problem. Many works have been conducted on spectrum sensing \cite{Chen, Pei,Ghasemi, Moayeri, LiFu, Beaulieu, Wang, Ma}. In \cite{Chen} and \cite{Pei}, spectrum sensing was studied without considering the primary user (PU) traffic during the sensing period. In \cite{Ghasemi, Moayeri,LiFu}, the authors considered the PU traffic in spectrum sensing but only assumed that the PU arrives at the channel before sensing starts. In \cite{Beaulieu, Wang, Ma}, a realistic scenario where the PU arrives or departs during the sensing period was discussed by assuming that the PU only changes its status once during the entire sensing period. Specifically, the effect of the PU traffic with one change on sensing performance was evaluated in \cite{Wang}. Better detectors for sensing based on the PU traffic were derived in \cite{Beaulieu} and \cite{Ma}. 

In this letter, the result in \cite{Wang} is generalized by considering a more practical case, where the PU changes its status an arbitrary number of times during the sensing period. This includes the scenario studied in \cite{Beaulieu, Wang, Ma} as a special case. It is also the case when the PU has high traffic or when a long sensing period is used. Different PU traffic models represented by different PU channel holding time distributions, including exponential \cite{Guerin, Bellamy}, log-normal \cite{Jedrzycki, Barcelo}, Gamma \cite{Li,Alwakeel} and Erlang \cite{Fang} are examined. The performance degradation caused by an arbitrary number of status changes is analyzed. Numerical results show that the status change of the channel during the sensing period causes performance degradation and the amount of degradation is related to the primary mean channel holding time. It is also shown that different holding time distributions cause different degradations.

%
%
%
%
%
%
%
%
%

\section{System Model And Derivation}
Consider a cognitive radio network where a licensed PU and an opportunistic secondary user (SU) operate in the same channel. A frame structure with duration $T$ is used by the SU, which consists of a sensing period $\tau$ and a transmission period $T-\tau$. Denote the total number of samples in the sensing period as $I$. The PU is assumed to arrive at or depart from the channel several times such that its status changes frequently during the sensing period. The number of PU status changes during the sensing period is an integer $x \in \left[0,N\right]$, where $N$ is the maximum number of changes with values from 1 to $I$, and $I \in [1,\infty )$. This takes an arbitrary number of status changes into account. Define the samples at which the PU arrives at and departs from the channel as $a_{j}$ and $d_{j}$, respectively, where $j$ represents the $j^{th}$ arrival or departure. 

The PU traffic is modeled as a 1-0 random process, where `1' represents PU presence and `0' represents PU absence. Denote the cumulative distribution functions (CDFs) of idle and busy durations of the channel as $F_{d}\left(t\right)$ and $F_{a}\left(t\right)$, respectively, which are determined by the PU traffic model and are assumed to follow any reasonable distributions. Since PU and SU are not synchronized and both PU and SU traffics are functions of time, there is always a PU status change that happened immediately before the current SU frame starts if one goes back in time from the instant when the SU frame starts, so that the same CDF applies to all the idle and busy durations in the SU frame regardless of when this status change occurs. This does not contradict with the exponential holding time and its memoryless property examined and used later. Assume that for all considered PU traffic models, each time interval can be modelled independently. In the case when the PU arrives at the channel at sample $a_j$ during the sensing period, the probability mass function (PMF) of $a_j$, $f_{a}\left(a_j\right)$ can be obtained from the CDF as \cite{Ma}
\begin{equation}  
f_{a}\left(a_j\right) = F_{a}\left(a_j t_s\right)-F_{a}\Big(\left(a_j-1\right)t_s\Big)
\end{equation}
where $t_s$ is the sample duration. Similarly, in the case when the PU departs from the channel during the sensing period at sample $d_j$, the PMF of $d_j$, $f_{d}\left(d_j\right)$ is derived as
\begin{equation}   
f_{d}\left(d_j\right) = F_{d}\left(d_j t_s\right)-F_{d}\Big(\left(d_j-1\right)t_s\Big).
\end{equation}

Denote the idle and busy channel hypotheses as $H_{x,0}$ and $H_{x,1}$, where $H_{x,0}$ represents the hypothesis that after $x$ status changes, the PU is absent at the end of the sensing period, and $H_{x,1}$ represents the hypothesis that after $x$ status changes, the PU is present at the end of the sensing period. The arrivals and departures occur one after the other. Thus, considering energy detection one has
\begin{equation}		\label{YH0}
Y_{H_{x,0}} = \begin{cases}
\begin{array}{llll}
\displaystyle\sum\limits_{j=1}^{\frac{x}{2}}\sum_{i=a_{j}+1}^{d_{j}}{\left(s_{i}+n_{i}\right)^2} + \displaystyle\sum\limits_{i=1}^{a_{1}}n_{i}^2 + \sum_{j=1}^{\frac{x}{2}-1}\sum_{i=d_{j}+1}^{a_{j+1}}{n_{i}^2} \\
\quad + \displaystyle\sum\limits_{i=d_{\frac{x}{2}}+1}^{I}{n_{i}^2}
, \quad\quad\quad\quad\quad\quad\quad\quad  x=even,\\
\displaystyle\sum_{j=1}^{\frac{x-1}{2}}\sum_{i=a_{j}+1}^{d_{j+1}}{\left(s_{i}+n_{i}\right)^2} + \displaystyle\sum\limits_{i=1}^{d_{1}}{\left(s_{i}+n_{i}\right)^2}\\
\quad +\displaystyle\sum\limits_{j=1}^{\frac{x-1}{2}}\sum_{i=d_{j}+1}^{a_{j}}{n_{i}^2} + \displaystyle\sum_{i=d_{\frac{x+1}{2}}+1}^{I} n_{i}^2 , \quad\quad x=odd
\end{array}
\end{cases}
\end{equation}
%
\begin{equation}		\label{YH1}
Y_{H_{x,1}}=\begin{cases}
\begin{array}{llll}
\displaystyle\sum\limits_{i=1}^{d_{1}}{\left(s_{i}+n_{i}\right)^2} + 
\displaystyle\sum_{j=1}^{\frac{x}{2}-1}\sum_{i=a_{j}+1}^{d_{j+1}}{\left(s_{i}+n_{i}\right)^2}+ \\
\quad \displaystyle\sum\limits_{i=a_{\frac{x}{2}}+1}^{I}{\left(s_{i}+n_{i}\right)^2} +
\displaystyle\sum\limits_{j=1}^{\frac{x}{2}}\sum_{i=d_{j}+1}^{a_{j}}n_{i}^2, x=even,\\
\displaystyle\sum_{j=1}^{\frac{x-1}{2}}\sum_{i=a_{j}+1}^{d_{j}}{\left(s_{i}+n_{i}\right)^2}+  \displaystyle\sum\limits_{i=a_{\frac{x+1}{2}}+1}^{I}{\left(s_{i}+n_{i}\right)^2}\\
\quad +\displaystyle\sum\limits_{i=1}^{a_{1}}{n_{i}^2}+
\displaystyle\sum\limits_{j=1}^{\frac{x-1}{2}} \sum_{i=d_{j}+1}^{a_{j+1}}{n_{i}^2},\quad \quad \quad \quad x = odd,
\end{array}
\end{cases}
\end{equation}
where $Y_{H_{x,0}}$ and $Y_{H_{x,1}}$ are outputs of the integrator under hypothesis $H_{x,0}$ and $H_{x,1}$, respectively, $n_{i}, \forall i\in\{1,\ldots,I\}$, are samples of the additive white Gaussian noise with mean zero and normalized variance one, and $s_{i}, \forall i\in\{1,\ldots,I\}$, are samples of the PU signal. 

\subsection{Probabilities of Occurring}
The PU changes occupancy status during the sensing period. At the beginning of the sensing period, the PU is present with probability $p_{b}$ and absent with probability $p_{e}=1-p_b$. In $H_{x,0}$, the PU is assumed to randomly arrive and depart for $x$ times before leaving the channel at the end of the sensing period. The conditional probability for $H_{x,0}$ can be derived as
\begin{equation}	
P_{H_{x,0}} = \begin{cases}
\begin{array}{ll}
p_{e}\displaystyle\prod\limits_{j=1}^{\frac{x}{2}}f_{a}\left(a_j\right)\prod^{\frac{x}{2}}_{j=1}f_{d}\left(d_j\right) \bigg(1-
\Big(F_{a}\left(It_{s}\right) \\
\quad\quad  -F_{a}\left(d_{\frac{x}{2}}t_{s} \right)\Big)\bigg), \quad \quad\quad \quad  x = even,\\
p_{b}\displaystyle\prod\limits_{j=1}^{\frac{x-1}{2}}f_{a}\left(a_j\right)\prod^{\frac{x+1}{2}}_{j=1}f_{d}\left(d_j\right)\bigg(1-\Big(F_{a}\left(It_{s}\right)\\
\quad \quad - F_{a}\left(d_{\frac{x+1}{2}}t_{s} \right)\Big)\bigg), \quad \quad \quad x=odd.
\end{array}
\end{cases}
\end{equation}
Note that when $x=0$, $P_{H_{x,0}}$ becomes $P_{H_{0,0}}= p_{e} \Big(1-F_{a}\left(It_{s}\right)\Big)$, similar to $H_0$ in \cite{Pei, Chen}. Derivations of (4) and (5) are given in \cite{17}.

In $H_{x,1}$, the PU is assumed to randomly arrive and depart for $x$ times before arriving at the channel at the end of the sensing period. The conditional probability of occurring for $H_{x,1}$ can be derived as 
\begin{equation} 
P_{H_{x,1}} = \begin{cases}
\begin{array}{ll}
p_{b}\displaystyle\prod\limits_{j=1}^{\frac{x}{2}}f_{d}\left(d_j\right)\prod^{\frac{x}{2}}_{j=1}f_{a}\left(a_j\right)\bigg(1- \Big(F_{d}\left(It_{s}\right) \\
\quad \quad - F_{d}\left(a_{\frac{x}{2}}t_{s} \right)\Big)\bigg),\quad \quad \quad x = even, \\
p_{e}\displaystyle\prod\limits_{j=1}^{\frac{x+1}{2}}f_{a}\left(a_j\right)\prod^{\frac{x-1}{2}}_{j=1}f_{d}\left(d_j\right)\bigg(1-
\Big(F_{d}\left(It_{s}\right) \\
\quad \quad - F_{d}\left(a_{\frac{x+1}{2}}t_{s} \right)\Big)\bigg),\quad \quad \quad x = odd.
\end{array}
\end{cases}
\end{equation}
Similarly, when $x=0$, $P_{H_{x,1}}$ becomes $P_{H_{0,1}} = p_{e}\Big(1- F_{d}\left(It_{s}\right) \Big)$, similar to $H_1$ in \cite{Pei,Chen}. Note that the above analysis applies to all channel holding time distributions due to the generality of the assumed CDF.

\subsection{Conditional Probabilities of False Alarm and Detection}
Next, we derive the conditional probabilities of false alarm and detection. During the sensing period, regardless of the number of status changes the PU has, at any time instant, the received sample has two possibilities: it either contains noise only, $n_{i}$, or contains signal and noise, $s_{i}+n_{i}$. 
Define the signal-to-noise ratio (SNR) of the received signal as $\gamma_p$. The conditional probability of false alarm can be derived as 
\begin{equation}	\label{Pfc}
P_{{fa}_{x,0}} = \frac{1}{2}erfc\left(\frac{\eta-E\left[Y\right]_{x,0}}{\sqrt{2 Var\left[Y\right]_{x,0}}}\right)
\end{equation}
where $erfc\left(\cdot\right)$ is the complementary error function, $\eta$ is the detection threshold, $E\left[Y\right]_{x,0}$ and $Var\left[Y\right]_{x,0}$ are the expectation and the variance of $Y_{H_{x,0}}$, respectively.
 
Define $Y_{1}=\sum n_{i}^2$, thus, according to the central limit theorem (CLT), the expectation and variance of $Y_{1}$ can be obtained as $E\left[Y_1\right] = k_1$ and $Var\left[Y_1\right] = 2k_1$, where $k_1$ is the total number of samples that contain only noise during the sensing period. Under hypothesis $H_{x,0}$,
\begin{equation}    
k_1=\begin{cases}
\begin{array}{llll}  
&I - \displaystyle\sum\limits^{\frac{x}{2}}_{j=1}d_{j} + \sum^{\frac{x}{2}}_{j=1}a_{j}, & \quad x = even,\\
&I - \displaystyle\sum\limits^{\frac{x+1}{2}}_{j=1}d_{j} + \sum^{\frac{x-1}{2}}_{j=1}a_{j}, & \quad x= odd.
\end{array}
\end{cases}
\end{equation}

Similarly, let $Y_2 = \sum \left(s_{i}+n_{i}\right)^2$, it can be derived that $E\left[Y_2\right]=k_2 + k_2 \gamma_p$ and $Var\left[Y_2\right]=2 k_2 + 4 k_2 \gamma_p$, where $k_2$ is the total number of samples that contain signal and noise during the sensing period. Under hypothesis $H_{x,0}$,
\begin{equation}   
k_2=\begin{cases}
\begin{array}{llll}
&\displaystyle\sum\limits_{j=1}^{\frac{x}{2}}d_{j}-\sum_{j=1}^{\frac{x}{2}}{a_{j}}, &\quad x = even,\\
&\displaystyle\sum\limits^{\frac{x+1}{2}}_{j=1}d_{j} - \sum^{\frac{x-1}{2}}_{j=1}a_{j}, &\quad x= odd.
\end{array}
\end{cases}
\end{equation}
Hence, it can be obtained that $E\left[Y\right]_{x,0} = I+k_2 \gamma_p$ and $Var\left[Y\right]_{x,0}=2I+4k_2 \gamma_p$ for (\ref{Pfc}).
\\*
%
The conditional probability of detection can be derived as 
\begin{equation}  \label{Pdc}
P_{d_{x,1}} = \frac{1}{2} erfc\left(\frac{\eta-E\left[Y\right]_{x,1}}{\sqrt{2 Var\left[Y\right]_{x,1}}}\right)
\end{equation}
where $E\left[Y\right]_{x,1}$ and $Var\left[Y\right]_{x,1}$ are the expectation and variance of $Y_{H_{x,1}}$ under hypothesis $H_{x,1}$. By using the same method as above, $E\left[Y\right]_{x,1}$ and $Var\left[Y\right]_{x,1}$ can be derived as
\begin{equation}  
E\left[Y\right]_{x,1} = \begin{cases}
\begin{array}{ll}
I+\left(I-\displaystyle\sum\limits_{j=1}^{\frac{x}{2}}a_j + \sum_{j=1}^{\frac{x}{2}}d_j\right)\gamma_p, & x=even,\\
I+\left(I-\displaystyle\sum\limits_{j=1}^{\frac{x+1}{2}}a_j + \sum_{j=1}^{\frac{x-1}{2}}d_j\right)\gamma_p, & x=odd
\end{array}
\end{cases}
\end{equation}

\begin{equation}    
Var\left[Y\right]_{x,1}=\begin{cases}
\begin{array}{ll}
2I+4\gamma_p\left(I-\displaystyle\sum\limits_{j=1}^{\frac{x}{2}}a_j + \sum_{j=1}^{\frac{x}{2}}d_j\right),\\
\quad \quad \quad \quad \quad \quad \quad \quad \quad \quad \quad \quad x=even,\\
2I+4\gamma_p\left(I-\displaystyle\sum\limits_{j=1}^{\frac{x+1}{2}}a_j + \sum_{j=1}^{\frac{x-1}{2}}d_j\right), \\
\quad \quad \quad \quad \quad \quad \quad \quad \quad \quad \quad \quad x=odd.
\end{array}
\end{cases}
\end{equation}

\subsection{Unconditional Probabilities of False Alarm and Detection}
According to the PU traffic model, the unconditional probability of false alarm can be obtained by averaging the conditional probability of false alarm in (\ref{Pfc}) over (5) as
\begin{equation}	\label{un-Pfa}
\overline{P_{fa}} = \frac{\displaystyle\sum\limits^{N}_{x=0} \Big(P_{fa_{x,0}} P_{H_{x,0}} \Big)}
{\displaystyle\sum\limits^{N}_{x=0} P_{H_{x,0}}}.
\end{equation}

Similarly, the unconditional probability of detection can be obtained by averaging (\ref{Pdc}) over (6) as
\begin{equation}	 \label{un-Pd}
\overline{P_{d}} = \frac{\displaystyle\sum\limits^{N}_{x=0} \Big(P_{d_{x,1}} P_{H_{x,1}}\Big)}
{\displaystyle\sum\limits^{N}_{x=0} P_{H_{x,1}}}.
\end{equation}

As an application of the above analysis, using a method similar to that in \cite{Pei}, the achievable throughput of the channel can be derived as
\begin{equation}	
\begin{array}{ll}
R\left(\tau\right) = 
&\displaystyle\sum\limits_{x=0}^{N}{P_{H_{x,0}}\left(1-\overline{P_{fa}}\right)}log_2\left(1+\gamma_s\right)\frac{T-\tau}{T} \\
&+\displaystyle\sum\limits_{x=0}^{N}{P_{H_{x,1}}\left(1-\overline{P_{d}}\right)}log_2\left(1+\frac{\gamma_s}{1+\gamma_p}\right)\frac{T-\tau}{T}.
\end{array}
\end{equation}
where $\gamma_s$ is the SNR for the secondary signal. Note that the above analysis is for one frame only, and an optimal value of $\tau$ is used which can be obtained by considering the sensing-throughput tradeoff discussed in \cite{Pei}. Due to the limited space, it is not shown here. When multiple frames are considered, our analysis will apply to the mega-frame that consists of these frames.

\section{Numerical results and Discussion}
In this section, the effect of PU traffic with multiple status changes during the sensing period is investigated by numerical examples. In all the examples, the Neyman-Pearson (NP) criterion is applied and the sample duration $t_s$ is set at 1 ms. The CLT still applies because it has been shown in \cite{Papoulis} that the sum distribution converges to Gaussian very quickly. Also, $N$ is the maximum number of changes while $x \in [0,N]$.

Fig. 1 examines the effect of PU traffic with multiple status changes on the sensing performance. Here, the sensing period $\tau$ is 20 ms, the received SNR is -5 dB, the exponential traffic model is applied with the mean busy holding time $\lambda_1^{-1} = 5$ ms and the mean idle holding time $\lambda_2^{-1} = 5$ ms. For illustration purpose and due to limited computing resources, a relatively small number of samples is used. As expected, when the number of primary status changes increases, the spectrum sensing performance degrades. However, the amount of degradation decreases as the number of status changes increases. This is because for the given values of $\tau$, $\lambda_1$ and $\lambda_2$ in this case, the probability that more than 4 status changes occur becomes very small. This suggests a minimum sensing performance due to the limited sensing period and mean holding time. Simulation results are also provided in Fig. 1. As can be seen, it agrees with the theoretical results very well.

Fig. 2 examines the effect of the PU traffic intensity on the sensing performance. Here, the channel mean holding time $\lambda_1^{-1} = \lambda_2^{-1} = 20$ ms. Comparing with Fig.1, one sees that, when $\lambda_1$ and $\lambda_2$ decreases, the sensing performance degradation becomes less significant for different values of $N$. This is because when $\lambda_1$ and $\lambda_2$ decreases, the channel mean holding time increases. The probability that the PU changes its status during the sensing period decreases. The sensing performance therefore is less affected by the PU traffic. 

Fig. 3 examines the effect of the received SNR on the sensing performance. In this case, the detection threshold is determined by letting $\overline{P_d}$ be 0.9, and the probability of false alarm $\overline{P_{fa}}$ is hence obtained by using the determined detection threshold. The exponential model is applied for the primary user traffic. As can be seen, the probability of false alarm increases when $N$ increases, indicating a degradation in the spectrum sensing performance when PU traffic with multiple status changes is considered. Moreover, although for all values of $N$, the spectrum sensing performance improves with the increase of the received SNR, this improvement becomes much smaller when the status change of the primary user traffic is taken into consideration. 

\begin{figure}[!t]
\centering
\includegraphics[width=2.9in]{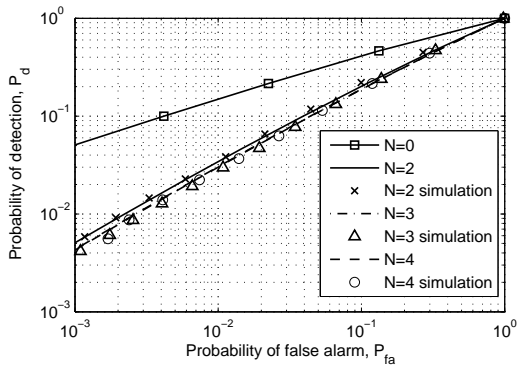}
\caption{The ROC for spectrum sensing with multiple PU status changes and exponential traffic for $\gamma_{p} = -5$ dB and $\lambda_1^{-1} = \lambda_2^{-1} = 5$ ms.}
\label{fig_sim1}
\end{figure}

\begin{figure}[!t]
\centering
\includegraphics[width=2.9in]{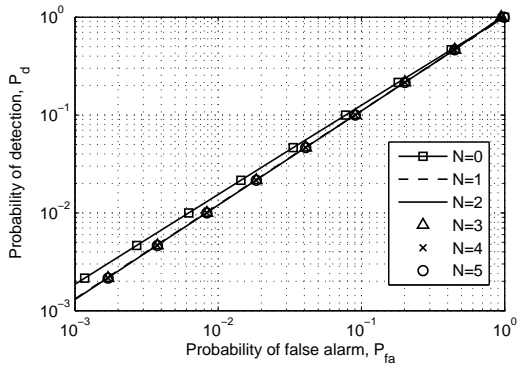}
\caption{The ROC for spectrum sensing with multiple PU status changes and exponential traffic for $\gamma_{p} = -15$ dB and $\lambda_1^{-1} = \lambda_2^{-1} = 20$ ms. }
\label{fig_sim2}
\end{figure}

Fig. 4 investigates the effect of PU traffic with multiple status changes on the sensing performance when different PU traffic models are applied. The mean channel holding time is set at 5 ms for all models. The maximum number of the PU status changes is $N = 5$. As can be seen, due to the difference in the traffic model, the PU with multiple status changes has different impacts on the sensing performance. In this case, the sensing performance is least sensitive to the Gamma model while it is most sensitive to the log-normal model. Similar patterns can also be found from the results of $N=1$ to $N=4$. This suggests that the same sensing scheme may have different performances when operating with different primary systems.

\begin{figure}[!t]
\centering
\includegraphics[width=3.1in]{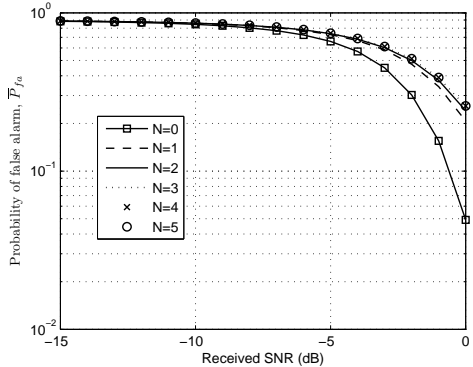}
\caption{The effect of PU traffic with multiple status changes for different received SNR.}
\label{fig_sim3}
\end{figure}

\begin{figure}[!t]
\centering
\includegraphics[width=2.9in]{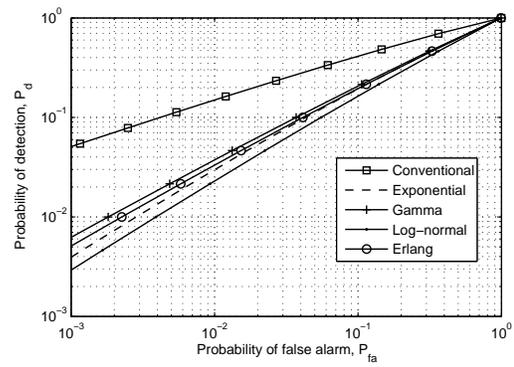}
\caption{The ROC for spectrum sensing with multiple PU status changes for different traffic models, when the mean holding time equals $5$ ms and $N=5$.}
\label{fig_sim5}
\end{figure}

\ifCLASSOPTIONcaptionsoff
\fi


\begin{thebibliography}{16}

\bibitem{Chen}
Y. Chen and N. C. Beaulieu, ``Performance of collaborative spectrum sensing for cognitive radio in the presence of Gaussian channel estimation errors'', \emph{IEEE Transactions on Commun.}, vol.57, no.7, pp. 1944-1947, Jul. 2009.

\bibitem{Pei}
Y. Pei, Y. -C. Liang, Kah C. Teh and Kwok H. Li, ``Sensing-throughput tradeoff for cognitive radio networks: A multiple-channel scenario,'' in \emph{IEEE 20th Int. Symp. on Personal, Indoor and Mobile Radio Commun., (PIMRC'09)}, pp. 1257-1261, Tokyo, Japan, Sept. 2009.

\bibitem{Ghasemi} 
A. Ghasemi and E. S. Sousa, ``Optimization of spectrum sensing for opportunistic spectrum access in cognitive radio networks'', \emph{IEEE 4th Conf. on Consumer Commun. and Networking, 2007, (CCNC 2007)}, pp. 1022-1026, Las Vegas, USA, 11-13 Jan. 2007. 
 
\bibitem{Moayeri}  
N. Moayeri and H. Guo, ``How often and how long should a cognitive radio sense the spectrum'', \emph{IEEE Symp. on New Frontiers in Dynamic Spectrum, 2010, (DySPAN 2010)}, pp. 1-10, Singapore, 6-9 Apr. 2010.

\bibitem{LiFu}
H. Li and H. Fu, ``An adaptive sensing period algorithm in cognitive radio networks'', \emph{IEEE Int. Conf. on Commun. Technology and Applications, 2009, (ICCTA'09)}, pp. 436-439, Beijing, China, 16-18 Oct. 2009.

\bibitem{Wang}
T. Wang, Y. Chen, E. L. Hines and B. Zhao, ``Analysis of effect of primary user traffic on spectrum sensing performance'', \emph{Int. Conf. on Commun. and Networking in China, 2009, (ChinaCOM 2009)}, pp. 1-5, Xian, China, 26-28 Aug. 2009. 

\bibitem{Beaulieu}
N. C. Beaulieu and Y. Chen, ``Improved energy detectors for cognitive radios with randomly arriving or departing primary users'', \emph{IEEE Signal Processing Letters}, vol. 17, issue. 10, pp. 867-870, Oct. 2010.

\bibitem{Ma}
J. Ma, X. Zhou and G. Y. Li, ``Probability-based periodic spectrum sensing during secondary communication'', \emph{IEEE Tran. on Commun.}, vol. 58, no. 4, pp. 1291-1301, Apr. 2010.

\bibitem{Guerin}
R. A. Guerin, ``Channel occupancy time distribution in a cellular radio system'', \emph{IEEE Trans. on Vehicular Technology}, vol. 36, no. 3, pp. 89-99, Aug. 1987.

\bibitem{Bellamy}
J. C. Bellamy, \emph{Digital Telephony}, 3rd ed. New York: John Wiley \& Sons. Inc., 2000.

\bibitem{Jedrzycki}
C. Jedrzycki and V. C. M. Leung, ``Probability distribution of channel holding time in cellular telephony systems'', \emph{IEEE Vehicular Technology Conf., 1996, (VTC 1996)}, vol. 1, pp. 247-251, Atlanta, USA, 28 Apr.- 01 May 1996.

\bibitem{Barcelo}
F. Barcelo and J. Jordan, ``Channel holding time distribution in cellular telephony'', \emph{IEEE Transactions on Vehicular Technology,}, vol. 49, no. 5, pp. 1615-1652, Sept. 2000.

\bibitem{Li}
X. Li and S. A. Zekavat, ``Traffic pattern prediction and performance investigation for cognitive radio systems'', \emph{IEEE Wireless Commun. and Networking Conf. 2008 (WCNC 2008)}, pp. 894-899,  Las Vegas, USA, 31 Mar.-03 Apr. 2008.

\bibitem{Alwakeel} 
M. Alwakeel, ``Deriving call holding time distribution in cellular network from empirical data'', \emph{Int. Journal of Computer Science and Network Security}, vol. 9, no. 11, pp. 93-95, Nov. 2009.

\bibitem{Fang} 
Y. Fang, I. Chlamtac and Y. -B. Lin, ``Modeling PCS networks under general call holding time and cell residence time distributions'', \emph{IEEE Transactions on Networking}, vol.5, no.6, pp. 893 - 906, Dec. 1997.

\bibitem{Papoulis}
A. Papoulis and S. U. Pillai, \emph{Probability, random variables and stochastic processes}, Fourth Edition, McGraw Hill, 2002.


\bibitem{17}
``Derivations of (4) and (5)'', http://www2.warwick.ac.uk/fac/sci/eng/ staff/yc/comlet\_derivations.pdf.
\end{thebibliography}
\end{document}